\documentclass{elsart}

\usepackage{natbib}

\usepackage{graphicx}
 
\begin{document}


\runauthor{Lara et al.}


\begin{frontmatter} 
\title{The Broad Line Radio Galaxy J2114+820} 
\author[IAA]{L. Lara\thanksref{TMR}},
\author[IAA]{ I. M\'arquez},
\author[NRAO]{W.D. Cotton},
\author[IRA]{L. Feretti}, 
\author[IRA,BOL]{G. Giovannini},
\author[UVAL]{J.M. Marcaide},
\author[IRA]{T. Venturi}

\thanks[TMR]{The author wishes to acknowledge that this research was 
partially supported by the European Commission's TMR Programme under contract 
No. ERBFMGECT950012}

\address[IAA]{Instituto de Astrof\'{\i}sica de Andaluc\'{\i}a, CSIC, Apdo. 3004, 18080 Granada, Spain} 
\address[NRAO]{NRAO, 520 Edgemont Road, Charlottesville, VA 22903-2475, USA}
\address[IRA]{Istituto di Radioastronomia, Via P. Gobetti 101, 40129 Bologna, Italy}
\address[BOL]{Dpto. di Fisica, Universit\'a di Bologna, Via Pichat 6/2, 40127 Bologna, Italy}
\address[UVAL]{Dpto. de Astronom\'{\i}a, Universitat de Val\`encia, 46100 Burjassot, Valencia, Spain}
 

\begin{abstract} 

In the frame of the study of a new  sample of large angular size radio
galaxies selected from  the NRAO  VLA Sky  Survey, we have  made  radio
observations of  J2114+820, a low power radio  galaxy with  an angular
size  of 6'.  Its  radio  structure basically consists  of a prominent
core,    a jet  directed   in north-west   direction  and two extended 
S-shaped
lobes.  We have also observed the  optical counterpart of J2114+820, a
bright elliptical galaxy  with a strong  unresolved central component.
The optical spectrum shows broad  emission lines.  This fact, together
with its low radio power and FR-I type morphology, renders J2114+820 a
non-trivial object  from the point of  view of the current unification
schemes of radio loud active galactic nuclei. 

\end{abstract} 



\end{frontmatter} 

\section{Introduction}
\label{intro} 

It is well known that Doppler beaming causes radio samples selected on
the basis  of a flux  density cutoff at high  frequencies to be biased
towards small orientation angles with
respect  to the  observer's line of  sight.  In order to properly test
unification schemes of radio loud  active galactic nuclei (AGN), it is
crucial     to  define samples   not    affected    by such bias    in
orientation. With this  in mind   among  other motivations, we   have
defined a new sample  of radio galaxies from the  NRAO VLA  Sky Survey
(NVSS) \citep{condon},  selecting   as    candidates those  features
apparently related as single objects, with emission over most of their
extent, and fulfilling the following restrictions: {\em i)} 
declination above +60$^{\circ}$; {\em ii)} angular size greater than 4';
{\em iii)} total flux density at 1.4 GHz greater than 100 mJy.

Candidates   have been   observed  with   the VLA  in   its  B  and  C
configurations at  1.4 and 5  GHz  in order to remove  possible random
coincidences of nearby  unrelated objects, but without overlooking  any
possible peculiar radio structure, and to  sort out the core emission
in  complex   structures, necessary   to  properly identify   optical
counterparts.   With work   almost  reaching  completion,  the  sample
comprises a total of 85 objects. A detailed description
of the sample will be presented elsewhere.

One of the  sources in our sample  is J2114+820, detected in the  NVSS
as  a 6' large radio source  with a total flux density
of 480 mJy.  Due to its prominent flat spectrum core, J2114+820 also 
belongs to  the 
Caltech - Jodrell
Bank  Flat-spectrum sample \citep{taylor}   and to  a  sample of  optically
bright flat spectrum sources  \citep{marcha}. It has a redshift of 0.084
\citep{stickel}. Its projected linear size is $\sim$500 Kpc and its total 
power P(1.4 GHz) = $6.7\times 10^{24}$   W/Hz (we assume H$_{0}$=75 km s$^{-1}$ Mpc$^{-1}$ and q$_{0}$=0.5).

\section{Radio Observations and Results} 
\label{radio} 

In the frame of the investigation of the radio galaxies in our sample,
we have made  VLA, MERLIN and  EVN observations  of J2114+820, mapping
its radio structure from kiloparsec to  parsec scales. 

\begin{figure}[pt] 
\centering
\includegraphics[scale=0.385,angle=0]{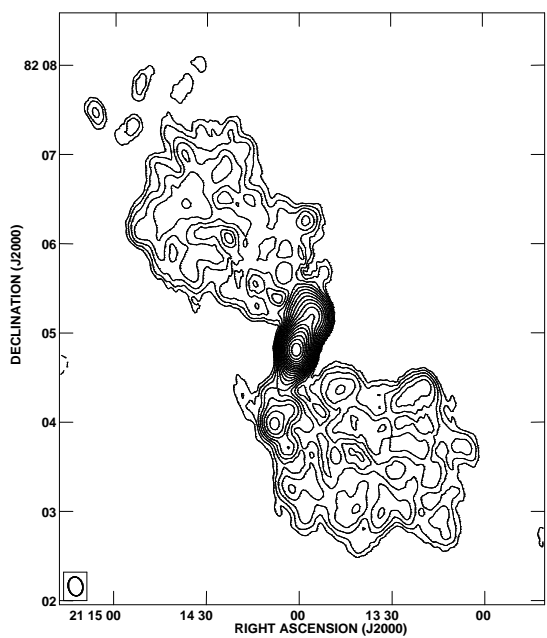}  
\caption{VLA map of J2114+820 at 1.4 GHz. A Gaussian beam of $12''.8  
\times 9''.7$  in P.A. 11$^{\circ}$ was used for
convolution. The lowest contour (at 3$\sigma$ of the rms of the image) is 0.369
mJy/beam. The peak of brightness is 174 mJy/beam. Contours increase in powers
of $\sqrt{2}$.}
\label{fig1} 
\end{figure}

In Figure  \ref{fig1}  we show  a VLA  map  of  J2114+820  at 1.4  GHz,
resulting from  the combination  of B and  C configuration   data obtained on
November 1995 and February 1996, respectively. The
radio structure is  dominated  by   a prominent and   unresolved
core. There is  a jet extending towards  the north-west,  and a weaker
counter-jet in opposite direction.  Both, jet and counter-jet, end  up in 
diffuse lobes of emission with  S-shaped  morphology. The  total power  and 
the radio structure  of
J2114+820 support its classification as a FR-I type radio galaxy. 

\begin{figure}[pb] 
\centering
\includegraphics[scale=0.45,angle=0]{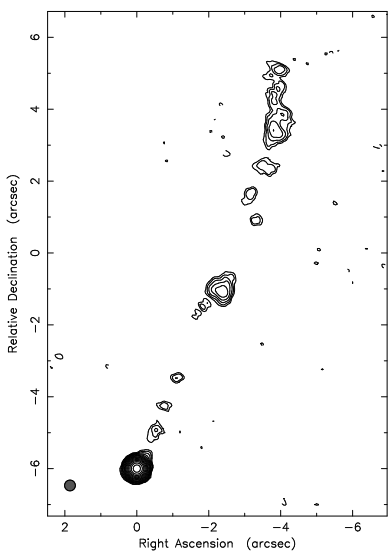}  
\caption{MERLIN map of J2114+820 at 1.66 GHz. A circular Gaussian beam 
of  $0''.3$ was used for
convolution. The lowest contour (at 3$\sigma$ of the rms of the image) is 0.355
mJy/beam. The peak of brightness is 142 mJy/beam. Contours increase in powers
of 1.5.}
\label{fig2} 
\end{figure}  

\begin{figure}[hbt] 
\centering
\includegraphics[scale=0.5,angle=0]{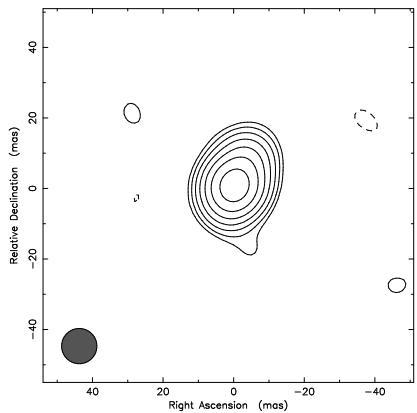}  
\caption{EVN map of J2114+820 at 1.66 GHz. A circular Gaussian beam of 10 mas
was used for
convolution. The lowest contour (at 3$\sigma$ of the rms of the image) is 0.92
mJy/beam. The peak of brightness is 91.5 mJy/beam. Contours increase in powers
of 2.}
\label{fig3} 
\end{figure}

The resulting map  after calibrating and imaging  1.66 GHz MERLIN data
obtained on June 1997 is displayed in Figure \ref{fig2}. Here again we
observe a strong  and unresolved core, with a  very faint and slightly
bent jet directed  towards the north-west. The  jet is not smooth, but
dominated by two  main components at distances  of $5''.6$ and $10''.5$ 
from  the core,  respectively.   Simultaneously with  MERLIN,  the EVN
observed  J2114+820  also  at 1.66  GHz,  with  antennas at Effelsberg
(Germany),  Simeiz (Ukraine),  Medicina and Noto  (Italy), Onsala (Sweden),
Westerbork (The Netherlands),  Torun  (Poland), Cambridge and  Jodrell
Bank  (U.K.).  The  EVN map,  in  Figure \ref{fig3},   shows a compact
structure  slightly elongated in the  direction of  the main jet. Data
can be fitted with two  Gaussian components (Table \ref{tab2}), in a manner 
fully consistent with results from   VLBA observations at   5      GHz
\citep{taylor}.  Combination of EVN and MERLIN data  does not add more
information since  the jet  appears  completely  
resolved  at the intermediate angular resolution provide by the combination 
of both interferometers.

\begin{table}[hb] 
\begin{center}
\caption{Gaussian fit parameters of the compact (EVN) structure of J2114+820}
\label{tab2}
\vspace{0.5cm}
\begin{tabular}{ccccccc}
\hline
Component &   S   &   D   &  P.A. &   L   &   r  & $\Phi$ \\
          & (mJy) & (mas) & (deg) & (mas) &      & (deg)  \\ \hline 
    1     & 80.6  &  --   &  --   &  4.8  &  0.1 & -14    \\
    2     & 34.9  &  5.1  & -23   &  9.3  &  0.3 & -30    \\
\hline \\
\end{tabular}   
\end{center}
\end{table}

\section{Optical Observations and Results} 
\label{optical} 

\begin{figure}[hbt] 
\centering
\includegraphics[scale=0.45,angle=0]{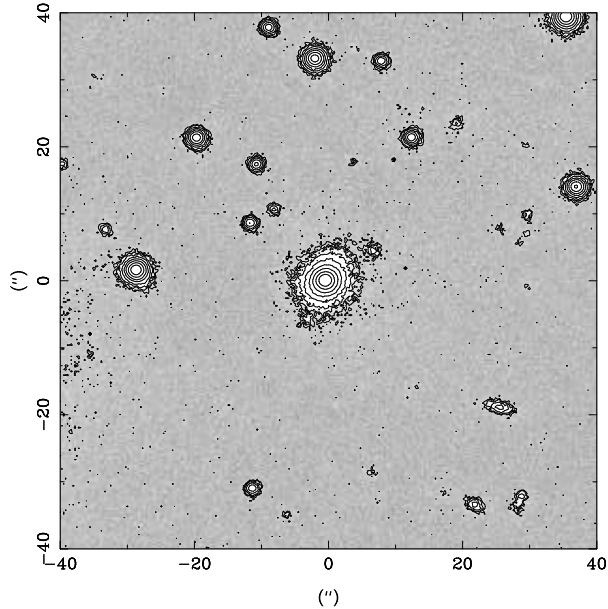}  
\caption{R-filter image of J2114+820 taken at the 2.2 m telescope in Calar Alto.}
\label{fig4} 
\end{figure}

We made observations of  the optical counterpart of J2114+820
with  the  2.2 m telescope at  the  Calar Alto  observatory (Spain) on
September 1997, making use of the direct imaging and spectroscopic 
capabilities of the Calar Alto Faint Object Spectrograph (CAFOS). 
The R-filter
image of  the field of J2114+820 after a 300 second exposure in Figure  
\ref{fig4} shows a
bright galaxy coincident with the  radio core position (the seeing is 
$\sim 1''$). 
The brightness
profile  of this galaxy is   consistent with that of an    elliptical 
(r$^{\frac{1}{4}}$) harboring an unresolved central component  associated 
with the active nucleus. There  are also hints of interaction  of this 
galaxy with one (or maybe two) dwarf galaxy in the surroundings. 
Atmospheric conditions prevented us from obtaining a reliable photometric 
calibration of our data.

\begin{figure}[hbt] 
\centering
\includegraphics[scale=0.5,angle=-90]{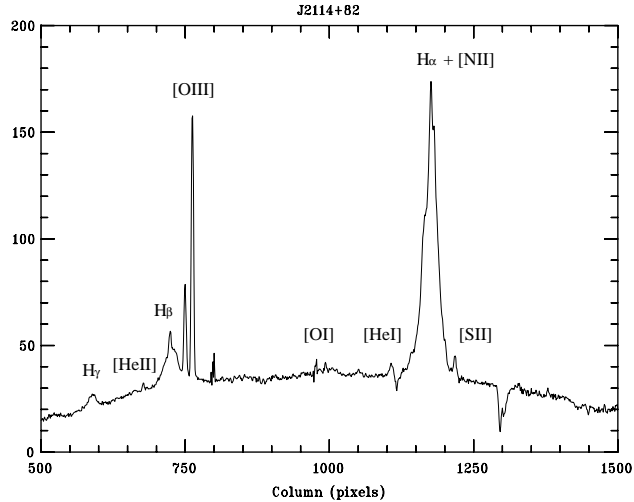}  
\caption{Spectrum of J2114+820.}
\label{fig5} 
\end{figure}

The spectrum of J2114+820 is displayed in Figure 
\ref{fig5}. We confirm the redshift and main features of the spectrum 
reported in  \citet{stickel}. J2114+820 exhibits strong broad Balmer lines,
but not only in the core region, but also at distances up to 4 Kpc from the 
center, indicating the existence of high velocity gas over large extensions 
of the galaxy.

\section{Discussion}
\label{tables} 

J2114+820 can be classified as a low power FR-I radio 
galaxy. According to current unification schemes of radio loud
AGN's  \citep{urry}, FR-I radio sources constitute the parent population of Bl-Lac type objects,
which are characterized by the absence of broad emission lines 
(at most, only weak broad lines have occasionally  been observed in Bl-Lac type objects \citep{vermeulen}). In 
consequence, FR-I sources should not show such emission lines, contrary 
to what we observe in J2114+820. In this sense, J2114+820 hardly fits the 
requirements of unification schemes. Another peculiarity of this galaxy is 
the evidence of high velocity gas at distances as large as 4 Kpc  from the 
core, a fact not common in elliptical galaxies which are relatively poor in 
gas.   
New observations are being analyzed in order to bring J2114+820 into a coherent
scenario.

\end{document}